\documentstyle{article}

\begin{document}
\centerline{MEANING OF THE DENSITY MATRIX}
\vskip 1cm
\centerline{~Y. Aharonov$^{*\dagger}$ ~and ~J. ~Anandan$^*$}
\centerline{*Department of Physics and Astronomy}
\centerline{University of South Carolina}
\centerline{Columbia SC 29208, USA}
\centerline{E-mail: jeeva@sc.edu}
\centerline{and}
\centerline{$^\dagger$ School of Physics, Tel Aviv University}
\centerline{Ramat Aviv, Tel Aviv, Israel}

\begin{abstract}

Protective measurement, which was proposed as a method of
observing the
wavefunction of a single system, is extended to the observation of
the density matrix of a single system. d'Espagnat's definition of
`proper mixture' is shown to be improper because it does not
allow for appropriate 
fluctuations. His claim that there could be different
mixtures corresponding to the same density matrix is critically
examined. These results provide a new
meaning to the density matrix, which gives it the same ontological
status as the wavefunction describing a pure state. This
also
enables quantum entropy to be associated with a single system.

\end{abstract}
\vskip 2cm
\noindent{December 26, 97, revised March 7, 98}


\newpage
\par

\newpage
\par

\section{Introduction}

It is well known that quantum mechanics may be formulated by
specifying the state of a quantum system by either a statevector
that belongs to the Hilbert space or by a `density matrix' that is a
Hermitian operator $\rho$ acting on the Hilbert space with
non negative eigenvalues whose sum is $1$. The density matrix
has the advantage that not only does it describe a pure state that
satisfies the condition $\rho^2=\rho$, which can equally well be
described by a state vector, but also it can describe a mixed
state which cannot be described by a state vector.

The traditional meaning given to the density matrix is that its
eigenvalues are the probabilities for finding the system in the
corresponding eigenstates. The only way we know of giving
physical meaning to the probabilities of outcomes is by
interpreting
them as relative frequencies of the corresponding states in an
appropriate ensemble of $N$ identical copies of the system that
are in the various possible states and taking the limit of
$N\rightarrow \infty$.

But there are three drawbacks to this meaning  due to the fact
that the ensemble associated with $\rho$ by this prescription is
not unique: (1) When two or more eigenvalues of $\rho$ are
equal, the basis of eigenstates of $\rho$ is not unique. (2) If a
usual measurement is made of an observable $A$ that does not
commute with $\rho$, then the possible outcomes, which are
eigenstates of $A$, are in general different from the eigenstates of
$\rho$. Hence $\rho$ undergoes a sudden change to a new density
matrix $\rho '$
whose eigenstates are the eigenstates of $A$. This is the quantum
measurement problem stated more generally than the usual
statement in which $\rho$ represents a pure state. Even if this
discontinuity in evolution is accepted, unless the eigenvalues of
$A$ are all distinct, the ensemble associated with $\rho '$,
according to the above prescription, is not unique. (3) As we shall
see later, the
interpretation of  the probabilities as relative frequencies is not
possible for a fixed finite $N$. We also cannot set $N=\infty$
because then the relative frequencies would be ratios of infinities
which are not mathematically meaningful. So, it is necessary to
take the limit of of $N\rightarrow \infty$. We then need to
associate with $\rho$ a sequence of finite ensembles with ever
increasing $N$, with none of the members of the sequence
providing a physical meaning to $\rho$, except in some
aproximate sense. It appears to us that this sequence is merely a
conceptualization and does not provide an objective reality that
could be described by the density matrix.

In this paper we provide a new meaning to the density matrix
which does not have the above drawabacks. We shall be
concerned, in this investigation, with the following questions: (A)
Does the density matrix of a system uniquely determine the state
of the
system?
(B) Can the density matrix be associated with a single system, as
opposed to an ensemble? We shall answer both these questions
affirmatively, and in the process correct some common
misconceptions concerning the density matrix. The new meaning
we shall give to the density matrix will be based primarily on
generalizing a method for observing a pure state by means of
`protective measurements' on a {\it single} system in that
state\cite{aaa} to the observation of the density matrix of a {\it
single} system. The philosophical aspects of this work will be
discussed elsewhere\cite{an1998}.

\section{Protective Measurements of the Density Matrix}

To fix our ideas, we shall consider just one way of doing protective
measurements: The state of the quantum system under
observation is in an eigenstate, described by the state vector
$|\psi >$, of the Hamiltonian $H$ with non degenerate eigenvalue.
An observable $A$ is measured adibatically in a time interval
$T>>{\hbar \over |\Delta E|}$, where $\Delta E$ is the energy
difference between the energy of the given state and the nearest
energy eigenvalue, so that there is no appreciable change in the
state.
This results in the observation of $<\psi |A|\psi >= tr~\rho A$
although only a single system is being used\cite{aaa}, where
$\rho=|\psi><\psi|$ is the density matrix of this pure state. By
doing protective observations of sufficient number of observables
$A$ the state vector $|\psi>$ may be determined up to phase.
Since the phase is undetermined, protective measurements really
determine the system's density matrix $\rho$ and not its state
vector.

In the above procedure it was assumed that the system was in a
non entangled state prior to the interaction with the apparatus.
Then the interaction with the apparatus does not lead to any
entanglement. But in an actual experiment, some interaction of the
 system with the environment is unavoidable. So, let us consider
the system under consideration and another system in an
entangled state. Then without loss of generality the state of the
combined system may be written as
\begin{equation}
|\chi> = \sum_i c_i |\psi_i>|\phi_i>
\end{equation}
where the $|\psi_i>$ and $\phi_j$ are, respectively, orthonormal
states of the first and the second system.

The protective measurement of a state of two systems,
represented by $|\chi>$,
has been studied by one of us\cite{an1993}. If the observable,
represented by the Hermitian operator $A$, that is so measured is
that of the first system only, then the result of the protective
measurement is
\begin{equation}
<\chi |A|\chi >=\sum_i |c_i|^2 <\psi_i |A|\psi_i >=tr\rho A
\end{equation}
where $\rho =\sum_i |c_i|^2 |\psi_i ><\psi_i |$ is the reduced
density matrix of the first system. Thus the possible pure states in
which the system can be in which are represented by $\rho$ all
contribute with the appropriate weights to the result of the
protective measurement which is a single number, such as the
displacement of the pointer in an apparatus. This is very different
from the protective observation of a single pure state which was
studied previously\cite{aaa,an1993}. We have obtained a new
physical meaning to $tr\rho A$ as a number that can in principle
be obtained in an appropriate single experiment, and not a mean
value of many experiments which was the original physical
meaning given to $tr\rho A$.

Protective measurements of different observables $A_\alpha$ of
the first system, gives $tr\rho A_\alpha$ for each of these
observables. By doing this for sufficient number of observables,
$\rho$ may be determined.  We emphasize that $\rho$ is obtained
in this way by
measurements on a single system, and not by measurements on
an ensemble of systems as done so far. This gives a new meaning
to the density matrix by answering question (B) in the
introduction affirmatively.

\section{Quantum Entropy}

There is an interesting difference between classical and quantum
entropy. Suppose we have a gas of classical molecules in a box. If
all the positions and velocities of the molecules are known then
the gas cannot have a non trivial entropy.
But suppose the box is now divided into many small cells and we
know only which cell each molecule is in. Similarly, the velocity of
each molecule is also known to some uncertainty. So, for each
microstate of the gas of molecules there is a unique
`coarse-grained' macrostate. The entropy of any microstate is
roughly the logarithm of the number of microstates which have
the same macrostate as the given microstate. Thus, classically,
entropy needs to defined with respect to a coarse-grained
observable
. Also, this entropy is associated with an ensemble of microstates.

But quantum mechanically it is not necessary to coarse-grain in
order to introduce the notion of entropy. The results of all the
measurements we make on a system are determined by its
density matrix $\rho$. If the measurements are of the usual kind,
then
these results are probabilitites and therefore need to be given
physical meaning by means of
a Gibbsian ensemble of identical copies of the given system. If the
measurements are protective then the results are definite values,
e.g. pointer readings of the
measuring apparatus, and so do not require an ensemble
interpretation. But in either case, $\rho$ completely determines
the results of measurements and therefore may be regarded as a
{\it complete} description of the system. Therefore, the quantum
entropy defined in terms of $\rho$ by
\begin{equation}
S=tr ~\rho ~ln\rho
\label{entropy}
\end{equation}
does not require any coarse-graining. Unlike the classical entropy,
quantum entropy is not defined with respect to any observable,
coarse-grained or otherwise. To sumarize, quantum entropy is
defined using only $\rho$, which has the maximum possible
information about
the system.

A consequence of this difference is that even if the box of gas
considered above is isolated from the environment, the classical
entropy would remain the same or increase whereas the quantum
entropy (\ref{entropy}) remains the same because $\rho$
undergoes unitary evolution.
However, if we divide the gas into subensembles then the
quantum entropy of each of them would in general increase as
these subensembles become more and more entangled as a result
of the interaction between the molecules. Thus quantum
 entropy is a measure of the degree of entanglement of the system
or the impurity of the density matrix, whereas classical entropy
has a very different meaning, namely it is a measure of the loss of
information. E.g. if the gas of molecules were intially
confined to a small part of the box, then it is overwhelmingly
likely that the gas would expand to fill the box . This is
accompanied by a corresponding increase in the classical entropy
which represents the decrease in information of the positions of
the
classical molecules. But the quantum entropy of an expanding, but
isolated, gas of quantum molecules remains the same because of
the unitary evolution.

However, the usual interpretation of quantum entropy, like
classical entropy, needs an ensemble of identical systems for its
physical meaning. Because $\rho$ can be determined by the usual
measurements only statistically and therefore this is equally true
 for the entropy (\ref{entropy}). Indeed, the above definition of
entropy without coarse-graining has been regarded by some as
possible because of the intrinsic statistical nature of quantum
theory, whereas in classical physics coarse-graining, or lack of
complete information, is needed to introduce the statistical
element.

But the new result obtained in section 2 is that $\rho$ may be
determined by protective measurements deterministically so that
$\rho$ may be associated with a single system. It then follows
that the quantum entropy (\ref{entropy}) also may be associated 
with a single system. In general, all
physical quantities computed using the density matrix will from
now on have a new meaning because of the new meaning to the
density matrix we have given by associating it with a single
system.

\section{Mixtures and Fluctuations}

We now turn to question (A) mentioned in the introduction,
namely whether the density matrix $\rho$ uniquely determines
the state of the system. As mentioned in the previous section,
$\rho$ gives a complete description of the state of the system.
Therefore, from an operational point of view, the state it describes
must be unique, because two states which have the same density
matrix cannot be experimentally distinguished. This conclusion
seems to differ from that of d'Espagnet\cite{esp}, whose
arguments therefore need to be considered here.

In arguing for the possibility of different types of mixtures with
the same density matrix, d'Espagnet makes a distinction between
`proper' and `improper' mixtures. He defines a proper mixture to
be a collection of $\mu$ ensembles $E_1, E_2,...E_\mu$ of physical
systems all of the same type with all the members of each
ensemble $E_\alpha$ being in the same state, described say by
the normalized state vector $|\phi_\alpha>$. He then regards the
operator
\begin{equation}
\rho = |\phi_\alpha>{N_\alpha\over N}<\phi_\alpha |
\label{dm}
\end{equation}
as the density matrix of this `mixture', where $N=\sum_\alpha
N_\alpha$ is the total number of elements in this collection.
An improper mixture is for him the state of a subsystem of a
larger system that is in a pure entangled state, which we
considered in section II.

However, the above mentioned collection does not qualify to be a
mixture because it does not allow for the appropriate 
fluctuations that a
genuine mixture ought to have, and the association of the above
density matrix with it is unjustified. To illustrate the first point we
shall consider the following two examples.

A famous cricketer who was leading his country's team in an
international cricket test series against another country 
lost the toss in the first four
tests. He then thought that according to the ``law of chances'' it is
overwhelmingly likely that he would win the fifth toss, even
though different coins were used for the different tosses, and so
packed his team with batsmen. He lost the fifth toss too. Even
more amusing is the possibly fictitious example of a man who was
found carrying a bomb in a plane. When arrested he defended
his action by saying ``I was not planning to explode this bomb. But
statisticians have told me that the probability of two passengers
carrying bombs in the same plane at the same time is much
smaller than the probability of just one passenger carrying a
bomb. I therefore wanted to decrease the probability of someone
else carrying a bomb on this plane by carrying one myself''.
The fallacy  common to the reasoning of both men is the failure to
recognize that the similar events whose probabilities are relevant
here are independent.

Another way of expressing this independence is terms of
fluctuations. If there are six tosses then the mean value of heads
coming up is three. But of course the remaining six possible values
of the number of heads can also occur, although with smaller
probabilities, and they consititute the fluctuations. They ensure
that even after the first four tosses turned up heads, there is an
equal probability for head or tail in the fifth toss. In general, the
fluctuations are such that the system has no memory. This
enables thermodynamics to be possible because a thermodynamic
state is described by variables such as pressure or temperature
which have no memory of the history of the system.

But the `proper mixture' as defined by d'Espagnet would have a
memory: After measurements are made on $n$ systems in the
mixture, the results obtained would be relevant in predicting the
probabilities of the states of the remaining $N-n$ systems,
because
the ratios of the numbers of systems in the various states for the
$N-n$ systems would in general be different from the ratios in the
originally chosen collection of $N$ systems. It is a memory such as
this one that the above mentioned cricket captain and the bomb 
carrying passenger were
vainly hoping for. It follows that we cannot represent the above
collection of d'Espagnet by the density matrix (\ref{dm}) because
the latter does not have this memory.

We can, however, improve on d'Espagnet's definition of a proper
mixture by taking the limit $N\rightarrow \infty$ while keeping
the ratios ${N_\alpha\over N}$ fixed. As $N$ becomes larger for a
fixed $n$, the memory we have from the first $n$ trials become
correspondingly more feeble. Then the density matrix (\ref{dm})
will legitimately represent the limit of this sequence of collections.
But it cannot be any one of this sequence, however large $N$ may
be. So, we  cannot point to the state of a physical
 system that is represented by $\rho$ in the sense that
measurements on this state would give $\rho$, as d'Espagnet was
trying to do, so long as only the usual measurements are
performed on the system. But by doing protective measurements
on a single system, as described above, we can determine $\rho$.
To be sure there is also an idealization in the above protective
measurements because they must be adiabatic. But the
experiments are performed on a single system which makes them
different from the usual statistical way of determining $\rho$ by
means of measurements on an ensemble of systems.

The second argument that d'Espagnet makes in support of the
view that the same density matrix can represent different
mixtures is the following: He states that it is possible to have
different mixtures with the same density matrix that can be
distinguished by their different methods
of preparation.
As an example he considers two methods of preparing an
unpolarized beam of $N$ spin $1/2$ particles, by mixing equal
amounts of particles with opposite spins with the quantization axis
along the $z-$ and $x-$ directions respectively
(ref. \cite{esp}, p. 121).
He claims that the density matrix is the same
in both  cases and it is
\begin{equation}
\rho= (1/2)I,
\label{d'e}
\end{equation}
where $I$ is the $2\times 2$ identity matrix.
He then shows correctly that the two ensembles can be
experimentally distinguished by the fact that the fluctuations for
the total spin in the $z-$direction $\Sigma_z =\sum_{n=1}^N
{\sigma}_{z,n}$ are different for them. Because the standard
deviations of this operator are $0$ and $\sqrt{N}$ for the first and
second ensembles respectively.

However, as already mentioned, the results of {\it all}
measurements on a quantum system can be obtained from its
density matrix. The measurement that d'Espagnet uses to
distinguish between the two ensembles is a measurement on the
entire ensemble and not the individual system. If the entire
ensemble is considered then the two ensembles are in different
pure states and therefore obviously the corresponding density
matrices are also different. So, it is not surprising that the two
ensembles can be experimentally distinguished. If the individual
system is considered then, as we have argued above, only in the
limit of $N\rightarrow \infty$ can the ensembles have the above
density matrix for the {\it spin} degrees of freedom. The
appropriate operator for the individual system is not $\Sigma_z$,
but rather the average spin in the $z-$ direction, namely
$(1/N)\Sigma_z$. The fluctuation of the latter operator is $1\over
\sqrt{N}$ which tends to zero as $N\rightarrow \infty$. So, in this
limit the two mixtures for the individual spin states cannot be
distinguished. This is consistent with the fact that both mixtures
have the same {\it spin} density matrix.

But what about d'Espagnet's claim that the two mixtures even
though they have the same density matrix can in some sense be
distinguished by their different methods of preparation? Actually,
careful examination of his example shows that when the two pairs
of beams in his example are combined, their density matrices will
be different if the position variables in addition to the spin
variables are taken into account. This is because of the spin
position correlation in the original beams and any method of
recombination must use a potential that varies with positon.
It is only after the position variables are traced over that the
reduced density matrix has the same form ( \ref{d'e} )  for the two
ensembles.

But it is of course possible to obtain the same final state density
matrix  starting from two different intial state density matrices by
using two different Hamiltonians. This does not mean that the
final states should be regarded as different. There will be no
disagreement with this statement if the density matrices all
describe pure states. E.g. a neutron spin state initially in the $X$ or
$Y$ direction may be rotated to the $Z$  direction by means of
magnetic fields in $Y$ or $-X$ directions, respectively. Although
the initial states were different, the final state is the same. So,
there is no reason why differently prepared states with the same
density matrix should not be regarded as the same state, if the
density matrix takes into account all variables.
Since there is no conceivable way of distinguishing between states
of a system with  the same density matrix by doing only
measurements on that system, we shall answer question (A) in the
introduction affirmatively.

In conclusion, we have shown here that the density matrix of a
mixed state is just as real as the density matrix of a pure state in
that they can both be associated with the state of a single system.
\vskip .5cm

{\bf ACKNOWLEDGEMENTS}

We thank Katherine Brading for drawing our attention to the
arguments in reference \cite{esp} considered in this paper
and for stimulating discussions. This work was supported
by NSF grant  PHY-9601280 and the work of Y.A. by the Israel 
National Academy of Sciences under grant no. 61495.

\end{document}